\documentclass[12pt]{article}

\begin{document}

\begin{center}

{\bf\large Seiberg's Duality from Monodromy of Conifold Singularity}

\vspace{3cm}

Kei Ito\\
Department of Electrical and Computer Engineering\\
Nagoya Institute of Technology\\
Gokiso, Showa-ku, Nagoya 466-8555, Japan

\vspace{1cm}
E-mail: keiito@eken.phys.nagoya-u.ac.jp or ito@ks.kyy.nitech.ac.jp

\vspace{5cm}
Abstract

\end{center}

Duality between N=1 supersymmetric gauge theories(Seiberg's
duality) is geometrized, in the framework of AdS/CFT correspondences.
It is shown that Seiberg's duality corresponds to monodromy
of wrapped D5 branes on the homology cycles of a generalized
conifold where D3 branes are located. The celebrated $\tilde
{N_{c}}=N_{f}-N_{c}, \tilde{N_{f}}=N_{f}$ rule is reproduced
and a braid group structure in a sequence of dualities, is found.

\vspace{1cm}

PACS; 11.25.Sq; 11.30.Pb

Keywords; Seiberg's duality, AdS/CFT correspondence, 
monodromy, conifold, singularity

\newpage

Recently, there has been great progress in our understanding
of the behavior of four dimensional supersymmetric gauge theories, 
in the strong coupling regime. In particular, Maldacena[1] conjectured
that N=4 supersymmetric gauge theory (conformal field theory)
in four dimensions, is dual to type IIB string theory on a 
product of five dimensional anti-de Sitter($AdS_5$) space and 
five dimensional sphere($S^5$). (AdS/CFT correspondence).
Subsequently, Klebanov and Witten[2] generalized AdS/CFT
correspondence to a supersymmetric gauge theory with less 
supersymmetry, i. e. N=1 . There they showed that a certain 
N=1 supersymmetric gauge theory, which arises as a field theory of
parallel threebranes near a conifold singularity of a six dimensional
Calabi-Yau space $Y_6$ corresponds to type IIB string theory on
$AdS_5 \times X_5$, with $X_5$ being
a five dimensional Einstein manifold, $(SU(2) \times SU(2))/U(1)$.
(See also ref.[3])
Gubser et al.[4] and Lopez[5] generalized this correspondences to a class
of more general conifolds $Y_6$ classified according to simply-laced
Lie algebras of ADE type, which is a cone over $X_5$ obtained by
blowing up an orbifold singularity of $S^5/ \Gamma $, with $\Gamma$
being diescrete group classified according to the same type 
of the simply-laced Lie algebra. These correspondences suggest that
the strong coupling behavior of N=1 supersymmetric gauge theories 
in four dimensions are encoded in geometry of corresponding conifold
singularity. There are dualities between N=1 supersymmetric gauge
theories(Seiberg's dualities)[6] where the dynamics in the strong
coupling regime are responsible. Therefore, it is tempting to 
conjecture that Seiberg's duality can be geometrized to an operation
on the conifold singularity by means of AdS/CFT correspondences. 
This is the problem which we address in this paper. We will find 
that Seiberg's duality corresponds to a geometrical operation on
the conifold which is a monodromy of homology 2-cycles of the 
conifold called Picard-Lefschetz monodromy. If there are D5 branes
wrapped on homology 2-cycles, in addition to a stack of D3 branes,
we will have a non-trivial Seiberg's duality and obtain the celebrated
rule, $ \tilde {N_{c}}=N_{f}-N_{c}, \tilde {N_{f}}=N_{f}$.
Moreover, we will find a braid group structure in a sequence of
Seiberg's dualities by virtue of this geometrization.

Following references [4],[5],we review N=1 supersymmetric field
theories which arise from D3 branes on generalized conifold 
of ADE-type. Consider N D3 branes located at an orbifold 
singularity $\bf{C} ^2/ \Gamma \times \bf{C}$, where $\Gamma$ is 
a discrete subgroup of SU(2) classified by simply-laced Lie algebra ADE.
The effective field theory on D3 branes is a four-dimensional
N=2 supersymmetric gauge theory with a product gauge group,
$\prod_{i=0}^{k} SU(Nn_i)$, where, k is the rank of the 
Lie algebra corresponding to the discrete group $\Gamma $,
and $n_i=dim \bf{r}_i$ with $\bf{r}_i$ being the i-th
representation of $\Gamma$. In the corresponding simply-laced
Lie algebra, $n_i$ is identical to Dynkin label of the node
of extended Dynkin diagram. The matters which arise are $a_{ij}$
copies of hypermultiplets in the bi-fundamental represenations,
$(n_iN, \overline{n_jN})$ where $a_{ij}$ comes from the 
decomposition $\bf{C}^2\otimes \bf{r}_i=\oplus_j \bf{C}^{a_{ij}}
\bf{r}_j$. The orbifold $\bf{C}^2/\Gamma $ can be represented
as an algebraic variety, by a hypersurface defined by the 
equation $f_{\Gamma }=0$ in $\bf{C}^3$ parametrized by three
complex numbers $x$,$y$ and $z$.
\begin{eqnarray}
f_{\Gamma}&=&x^{k+1}+y^2+z^2, \hspace{1cm} (for A_k)\\
f_{\Gamma}&=&x^{k-1}+xy^2+z^2, \hspace{1cm} (for D_k)\\
f_{\Gamma}&=&x^4+y^3+z^2, \hspace{1cm} (for E_6)\\
f_{\Gamma}&=&x^3y+y^3+z^2, \hspace{1cm} (for E_7)\\
f_{\Gamma}&=&x^5+y^3+z^2. \hspace{1cm} (for E_8)
\end{eqnarray}
The equation $f_{\Gamma}=0$ is invariant under the 
$\bf{C}^{\ast}$ action, $x\rightarrow \lambda^{\alpha}x,
y\rightarrow \lambda^{\beta}y, z\rightarrow \lambda^{\gamma}z$
where, $\alpha =1, \beta =\frac{k+1}{2}, \gamma =\frac{k+1}{2}$,
in $A_k$ case,for instance, since $f_{\Gamma}$ scales as 
$\lambda ^h f_{\Gamma}$, with$h=k+1$, under this action.

An N=1 supersymmetric theory is obtained if the orbifold
singularity is resolved to become a non-compact Calabi-
Yau manifold $Y_{\Gamma}$ with a conifold-like singularity.
The orbifold singularity is deformed by parameters t's, 
which is then fibered on a complex plane parametrized by
$\phi $. The defining equation of this ``generalized conifold''
is then given by;
\begin{equation}
F_{\Gamma}(\phi ,x, y, z)=\phi ^h f_{\Gamma}
(\frac{x}{\phi ^\alpha}, \frac{y}{\phi ^\beta} ; t)+z^2=0.
\end{equation}
For instance, the holomorphic function F for $\Gamma =A_k$ is;
\begin{eqnarray}
F_{\Gamma}(\phi ,x, y, z)&=&x^{k+1}+\phi ^2 t_2 x^{k-1}+...+\phi ^k
t_k x+y^2+z^2\\
&=&\prod_{i=1}^{k+1} (x-\xi _i\phi )+y^2+z^2
\end{eqnarray}
Let us find the effective field theory on the branes located
at this generalized conifold. The T-dual brane configuration 
pictures[7],[8],[9] are more convenient for this purpose. 
Let N D3 branes (world-volume in 0123 directions) be located
at the orbifold singularity of ALE space ($A_k$-type, for
simplicity)with coordinates 6789. Perform T-duality along
6ith direction, corresponding to the U(1) isometry of
the ALE space.[7] Then we obtain type II A string with k+1
NS 5 branes (world-volume in 012345 directions). D3 branes
are transformed into D4 branes (01236).The effective field
theory on the D4 branes, is N=2 supersymmetric gauge theory 
with the gauge group $SU(N)^{k+1}$. If, N D3 branes are located 
at a generalized conifold,[8],[9] instead of ALE[7], we obtain, 
k+1 NS 5 branes rotated in the complex surface spanned by, $x=x^8+ix^9$
and $\phi =x^4+ix^5$. In this case, scalar fields in adjoint 
representations of $SU(N)^{k+1}$ become massive and supersymmetry
is broken from N=2 to N=1.
The spacing of i-th and (i+1)-th NS 5 branes in 6-th direction,
is related to the coupling constant $g_i$ of the i-th factor
$SU(N)_i$ of the product gauge group, by;
\begin{equation}
\frac{x^6(NS_{i+1})-x^6(NS_i)}{g_s}=\frac{1}{g_i^2}
\end{equation}
where $g_s$ is the string coupling constant.
The spacing in 7-th direction, on the other hand, is related 
to the coefficient of Fayet-Iliopoulos(FI) term $\zeta _i$
of the i-th gauge factor through,
\begin{equation}
x^7(NS_{i+1})-x^7(NS_i)=\zeta _i.
\end{equation}
Seiberg's duality with respect to the i-th gauge group factor
of the product group corresponds to an interchange of i-th and
(i+1)-th NS 5 branes by the following procedure.[8],[10],[11]
The i-th and the (i+1)-th NS 5 branes come close together
in 6-th direction until $x^6(NS_{i+1})-x^6(NS_i)$ reaches
a small real value $\delta$.
In the field theory side, the coupling costant $g_i$ becomes
larger and larger until $\frac{1}{g_i^2}$ reaches $\frac{\delta}
{g_s}$. Then the FI term $\zeta _i$ turns on in order to avoid
passing through the infinite coupling constant point
$g_i=\infty$. Then $g_i$ and $\zeta_ i$ vary as;
\begin{equation}
\frac{g_s}{g_i^2}+i\zeta _i=\delta e^{\pi it}.
\end{equation}
where, t runs from 0 to 1. At t=1, FI term is turned off and
the order of two NS5 branes are interchanged in 6-th direction,
since 6-th and 7-th coordinates of NS5 branes change as;
\begin{equation}
[x^6(NS_{i+1})-x^6(NS_i)]+i[x^7(NS_{i+1})-x^7(NS_i)]=\delta
e^{\pi it}.
\end{equation}
Finally the separation of NS5 branes becomes larger and
larger until the gauge coupling constant comes back
to the original value.

Now let us come back to the original picture, i.e. type II B
string with N D3 branes at the conifold, by T-duality 
transformation. Then, $x^6(NS_{i+1})-x^6(NS_i)$ and 
$x^7(NS_{i+1})-x^7(NS_i)$ are real and imaginary part
of complexified k\"{a}hler class[k] of homology 2-cycles of
the conifold. We find that
the procedure corresponding to Seiberg's duality;
two NS5 branes come clos together, being interchanged,
and go away, is a geometrical operation
on the singularity of the conifold, which is called
``Picard-Lefschetz monodromy''.

Picard-Lefschetz monodromy of a singularity is defined as follows.
See the book by Arnold et al.[12] for detail. This monodromy
plays also a crucial role in two-dimensional conformal field theory[13]. 
Consider generalized conifold of $A_k$-type defined
by the equation;
\begin{equation}
F_{\Gamma }(\phi , x, y, z)=0
\end{equation}

The level manifold of level $\lambda \in \bf{C}$, is defined by the
equation, $F_{\Gamma }(\phi , x, y, z)=\lambda$.
The critical points of the holomorphic function $F_{\Gamma }$ is defined
by, 
\begin{equation}
\frac{\partial F_{\Gamma }}{\partial x} =0, \frac{\partial F_{\Gamma}}
{\partial y} =0,
\frac{\partial F_{\Gamma }}{\partial z} =0.
\end{equation}

This is a k-th order algebraic equation which has k solutions;
$x=X_i (i=1,2,...,k), y=0, z=0$. The values of the holomorphic
function F at critical points $x=X_i$, are called critical
values $\lambda _i (i=1,...,k)$. For example, for $\Gamma =A_k$
\begin{equation}
\lambda _i=\prod_{j=1}^{k+1} (X_i-\xi _j\phi).
\end{equation}
If the level value $\lambda$ starts from a non-critical value
$\lambda _0$, goes around one of the critical values $\lambda _i$
and comes back to the the original value $\lambda _0$, the level
manifold comes back to the original one but the homology cycles
(vanishing cycles) are reshuffled. The holomorphic function
F is expanded around the critical values;
\begin{equation}
F_{\Gamma }=\lambda _i+w^2_1+w^2_2+w^2_3.
\end{equation}
Then, i-th vanishing cycle 
$\Delta _i$ is[12]
\begin{equation}  
\Delta _i=\sqrt{\lambda -\lambda _i}S^2,
\end{equation}
where, $\sqrt{\lambda -
\lambda _i}$ is the complexified kahler class and $S^2$ is
a unit two-sphere.
\begin{equation}
S^2=\{(w_1, w_2, w_3) \mid \sum_{j=1}^{3} w_j^2=1, Im w_j=0\}
\end{equation}
The Picard-Lefschetz monodromy is the operation where the level 
value $\lambda$ goes around a critical value $\lambda _i$.
\begin{equation}
\lambda -\lambda _i=\delta^2e^{2\pi it}, for small \delta,
t\in [0, 1].
\end{equation}
This implies that 
\begin{equation}
\sqrt{\lambda -\lambda _i}=\delta e^{\pi it}
\end{equation}
Since $\sqrt{\lambda -\lambda _i}$ is the complex kahler class
[k] of the i-th vanishing cycle.
\begin{equation}
\sqrt{\lambda -\lambda _i} =[k].
\end{equation}
the change of complex kahler class [k] under the
Picard-Lefschetz monodromy is the same as that under 
the interchange of 
the i-th and (i+1)-th NS 5branes in the T-dual picture,
which corresponds to Seiberg's duality. Thus we find that Seiberg's
duality corresponds to Picard-Lefschetz monodromy of vanishing 
cycles of the corresponding conifold.

The gauge theory arising from N D3 branes at generalized conifold
is self-dual under the Seiberg's duality transformation. This
implies that the number of colors of the dual theory are the same
as those of the original theory. A non-trivial Seiberg's duality
where the dual gauge group is different from the original one
is realized when D5 branes are wrapped on the vanishing cycles
(homology cycles) of the conifold. The wrapped D5 brane[14] has D3 brane
charge which is the same as the wrapping number. Therefore, if a D5
brane is wrapped on th i-th vanishing cycle $\Delta _i$, with
wrapping number $M_i$, the effective D3 brane number is changed from N 
to $N+M_i$. Then the i-th gauge group factor is changed from 
$SU(N)_i$ to $SU(N+M_i)_i$. Under the Picard-Lefschetz monodromy 
transformation, the vanishing cycles are reshuffled, and wrapping
numbers of D5 branes are changed accordingly. In the field theory
side, this implies that the numberof colors is changed under Seiberg's
duality transformation.

Let us consider a simple example; the generalized conifold
of $A_3$ type. N D3 branes on this conifold give N=1 supersymmetric
gauge theory with the gauge group $G=SU(N)_0\times SU(N)_1 \times
SU(N)_2 \times SU(N)_3$. Each gauge group factor corresponds to
the node of extended Dynkin diagram of the Lie algebra $A_3$.
The generalized conifold of $A_3$ type has three vanishing cycles
(homology 2-cycles) $\Delta_1, \Delta_2$ and $\Delta_3$, corresponding
to $SU(N)_1, SU(N)_2$ and $SU(N)_3$. If D5 branes wrap around 
$\Delta_1, \Delta_2$ and $\Delta_3$, with wrapping numbers $M_1,
M_2$ and $M_3$, the gauge group is,
\begin{equation}
G=SU(N)_0\times SU(N+M_1)_1\times SU(N+M_2)_2\times SU(N+M_3)_3
\end{equation}
There are three Picard-Lefschetz monodromy transformations
$h_1, h_2$ and $h_3$. The transformation rules of vanishing
cycles $\Delta _i, (i=1,2,3)$ under $h_i, (i=1,2,3)$ are
calculated by use of Picard-Lefschetz formula[12].

\begin{equation}
h_i(a)=a+(a\circ \Delta _i)\Delta _i
\end{equation}
where intersections of two vanishing cycles $\Delta _i$
and $\Delta _j$ are represented by the intersection matrix
$S_{ij}=\Delta _i\circ \Delta _j=-A_{ij}$ where $A_{ij}$
is the Cartan matrix of $A_3$. Hence,

\[
S_{ij}=
\left(
\begin{array}{rrr}
-2 & 1 & 0\\
1 & -2 &1\\
0 & 1 & -2
\end{array}
\right)
\]

The action of $h_i(i=1,2,3)$ on $\Delta _j(j=1,2,3)$ can be
represented by the monodromy matrix defined by the relation;
\begin{equation}
h_i(\Delta _j)=(h_i)_{jk}\Delta _k
\end{equation}

Then,
\[
h_1=
\left(
\begin{array}{rrr}
-1 & 0 & 0\\
1 & 1 & 0\\
0 & 0 & 1
\end{array}
\right)
\]
\[
h_2=
\left(
\begin{array}{rrr}
1 & 1 & 0\\
0 & -1 & 0\\
0 & 1 & 1
\end{array}
\right)
\]
\[
h_3=
\left(
\begin{array}{rrr}
1 & 0 & 0\\
0 & 1 & 1\\
0 & 0 & -1
\end{array}
\right)
\]  
 
Now Seiberg's duality transformation performed on i-th
gauge group factor $SU(N+M_i)_i$ corresponds to i-th 
Picard-Lefschetz monodromy transformation, represented 
by i-th monodromy matrix $h_i$. By this matrix, the 
vanishing cycles are transformed as;
\begin{equation}
\Delta '_j=(h_i)_{jk}\Delta _k.
\end{equation}
Then the wrapping numbers of D5 branes are changed accordingly;
\begin{equation}
M_j\Delta _j\rightarrow M_j\Delta '_j=M_j(h_i)_{jk}
\Delta_k=M'_k\Delta _k 
\end{equation}
Hence;
\begin{equation}
M'_k=M_j(h_i)_{jk}
\end{equation}

If Seiberg's duality transformation is performed on the 
gauge group factor $SU(N+M_1)_1$, the corresponding monodromy
matrix is $h_1$. Hence,
\begin{equation}
M'_1=M_j(h_1)_{j1}=M_2-M_1,
M'_2=M_j(h_1)_{j2}=M_2,
M'_3=M_j(h_1)_{j3}=M_3.
\end{equation}

and the dual group turns out to be;
\begin{equation}
\tilde G =SU(N)_0\times SU(N+M_2-M_1)_1\times
SU(N+M_2)_2\times SU(N+M_3)_3.
\end{equation}

Under this duality transformation, coupling constant of
the gauge group factor with the suffix 1, becomes large,
wich should then be regaded as the color group. therefore,
the numbers of colors and flavors of the original theory are,
$N_c=N+M_1$ and $N_f=2N+M_2$, respectively,and those of the
dual theory are $\tilde N_c=N+M-2-M_1$ and $\tilde N_f=
2N+M_2$, respectively.

Note that the rule, $\tilde N_c=N_f-N_c$, $\tilde N_f=N_f$ is 
satisfied in this example. 

Seiberg's dual on group factor with the suffices 2 and 3 can
be taken in the same fashion, and the above rule is satisfied
also in these cases.

A general expression for the dual gauge theory can now be written.
Let the N=1 supersymmetric gauge theory with the gauge group
$G=SU(N)_0\times SU(n_1 N+M_1)_1\times...\times SU(n_kN+M_k)_k$
be the field theory which arises from N D3 branes on the conifold
given by a below up of $\bf{C}^2/ \Gamma \times \bf{C}$, with
$ \Gamma$ being
a discrete subgroup corresponding to a rank k simply-laced Lie
algebra, with Dynkin labels $n_i$, and D5 branes wrap around
i-th(i=1,...,k) homology 2-cycles with wrapping number$ M_i$.
The matters arise as hypermultiplets in the bi-fundamental
representations $(n_iN+M_i, \overline{n_jN+M_j})$, if i-th and 
j-th gauge group factors are ``adjacent'' in the the sense
that the nodes are connected by spokes in the corresponding
extended Dynkin diagram. Then the dual theory is determined 
as follows. The intersection matrix of homology 2-cycles of 
the conifold is;

\begin{equation}
S_{ij}=\Delta _i\circ \Delta_j =-A_{ij}
\end{equation}

where $A_{ij}$ is the Cartan matrix of the Lie algebra.
Then the i-th monodromy matrix $h_i$ can be calculated
by Picard-Lefschetz theorem;

\begin{equation}
h_i(a)=a+(a \circ \Delta_i) \Delta_i.
\end{equation}

Now if we perform Seiberg's duality transformation on i-th
gauge group factor, the homology 2-cycles $\Delta_i$ are transformed
by i-th monodromy matrix $h_i$. Then the dual gauge group $\tilde G$
is,

\begin{equation}
\tilde G =SU(N)_0\times SU(n_1N+M'_1)_1\times ...\times SU(n_kN+
M'_k)_k
\end{equation}

where

$M'_\alpha =M_\beta (h_i)_{\beta \alpha},and  
(h_i)_{\beta \alpha}: \beta \alpha$ component of i-th monodromy 
matrix.
The dual matters arise as hypermutiplets in the bi-fundamental
representations $(n_iN+M'_i, \overline{n_jN+M'_j})$ if i-th
and j-th gauge group factors are adjacent in the sense that
the nodes are connected by spokes in the corresponding extended
Dynkin diagram.

The geometrization of Seiberg's duality given above
provides us with a deep insight in properties of the duality.
We find a new structure in a sequence of Seiberg's dualities,
i.e. the ``braid group'' structure. It is known in mathematics[12]
and confirmed easily in simple examples, that the monodromy
matrices $h_i$ satisfy the following relations;

\begin{equation}
h_{i+1} h_i h_{i+1}=h_i h_{i+1} h_i
\end{equation}
\begin{equation}
h_i h_j=h_j h_i
\end{equation}
  
for
\begin{equation}
\mid i-j \mid \geq 2
\end{equation}

This is the algebra of ``braid group'' with k strands. This
fact combined with the relation between Seiberg's duality
and Picard-Lefschetz monodromy, shows that a sequence of 
Seiberg's dualities taken on the gauge group factors of
the product group must have the same ``braid group'' structure.
In particular, we find from the relation of the braid algebra
that successive Seiberg's duality transformations performed on
an adjacent gauge group factors do not commute, whereas those
on non-adjacent group factors commute. For example, let the
gauge group of the original theory be;
\begin{equation}
SU(N)_0\times SU(N+M_1)_1\times SU(N+M_2)_2\times SU(N+M_3)_3.
\end{equation}
From the braid algebra, it follows that the Seiberg's
duality performed on the group factor $SU(N+M_1)_1$ first
and then on $SU(N+M_2)_2$, and the Seiberg's duality 
performed in the reversed order, i.e. on the gauge group
factor $SU(N+M_2)_2$ first and then on $SU(N+M_1)_1$, give 
different theories. Because, in the former case, the 
resulting group is

\begin{equation} 
SU(N)_0\times SU(N+M'_1)_1\times SU(N+M'_2)_2\times SU(N+M'_3)_3
\end{equation}

with,
\begin{equation}
M'_i=M_k(h_1)_{kj}(h_2)_{ji}
\end{equation}
whereas in the latter case, the resulting group is,
\begin{equation}
M'_i=M_k(h_2)_{kj}(h_1)_{ji}.
\end{equation}

The resulting groups are different in two cases since 
$h_1 h_2\neq h_2 h_1$

However, if we perform Seiberg's duality transformation
three times succesively, in the distinct two ways given
below, we reach the same theory.

1. Perform Seiberg's duality on $SU(N+M_1)_1$ first, then
on $SU(N+M_2)_2$ and finally on $SU(N+M_1)_1$.

2. Perform Seiberg's duality on $SU(N+M_2)_2$ first, then 
on $SU(N+M_1)_1$ and finally on $SU(N+M_2)_2$.

This fact follows from the relation,
\begin{equation}
h_1 h_2 h_1=h_2 h_1 h_2
\end{equation}

In conclusion, we have geometrized dualities between N=1 
supersymmetric gauge theories in four dimensions (Seiberg's
duality), by means of AdS/CFT correspondence. It turned out
that Seiberg's duality can be interpreted as a geometrical
operation on the conifold, called Picard-Lefschetz monodromy.
The homology 2-cycles of the conifold are transformed by this
monodromy operation. If D5 branes wrap around these 2-cycles,
and have D3 brane charges corresponding to the wrapping numbers,
the effective D3 brane numbers are changed, by the monodromy
transformation. In the field theory side, this implies that the
number of colors changes upon Seiberg's duality transformation
and it turned out that the rule
$\tilde{N_{c}} = N_{f}-N_{c},\tilde{N_{f}} = N_{f}$ 
is satisfied. Furthermore, we obtained
a general expression for the dual gauge group, when Seiberg's
duality is taken on one of the gauge group factor of the 
product gauge group of the N=1 supersymmetric gauge theory
arising from the field theory on D3 branes on generalized
conifold singularity of ADE type. Moreover, we found a braid
group structure in a sequence of Seiberg's dualities taken
on the gauge group factors of the product gauge group, by
virtue of the geometrization of Seiberg's duality in the 
framework of AdS/CFT correspondences.

\begin{table}

\end{table}
\newpage
References

[1]J. Maldacena, Adv. Theor. Math. Phys, 2(1998)231

[2]I. R. Klebanov and E. Witten, hep-th/9807080

[3] D. R. Morrison and M. R. Plesser, hep-th/9810201

[4]S. Gubser, N. Nekrasov and S. Shatashvili, hep-th/9811230

[5]E. Lopez, hepth/981202

[6]N. Seiberg, Nucl. Phys. B 435(1995)129

[7]Kei Ito, hep-th/9712225

[8]A. M. Uranga, hep-th/9811004

[9]K. Dasgupta and S. Mukhi, hep-th/9811139

[10]R. von Unge, hep-th/9901091

[11]S. Elitzur, A. Giveon and D. Kutasov, Phys. Lett. B 400(1997)269

[12]V. I. Arnold, S. M. Gusein-Zade and A. N. Varchenko,
 
Singularities of Differential Maps'' vol I, II. Birkh\"{a}user,

1988, Boston, Basel, Berlin.

[13]Kei Ito, Phys. Lett. B 229(1989)379

[14]D-E. Diaconescu, M. R. Douglas and J. Gomis, hep-th/9712230

\end{document}